\documentstyle[preprint,aps,prd,epsfig]{revtex}
\tightenlines
\begin{document}
\draft
\newcommand{\eqa}{\begin{eqnarray}}
\newcommand{\ena}{\end{eqnarray}}
\newcommand{\eq}{\begin{equation}}
\newcommand{\en}{\end{equation}}
\newcommand{\bfalpha}{{\mbox{{\boldmath$\alpha$}}}}
\newcommand{\bfsigma}{{\mbox{{\boldmath$\sigma$}}}}
\newcommand{\bfgamma}{{\mbox{{\boldmath$\gamma$}}}}

\title{
Bound $q\bar{q}$ Systems in the Framework of the
Different Versions of the 3-Dimensional
Reductions of the Bethe-Salpeter Equation
}

\author{
T.~Babutsidze$^1$, T.~Kopaleishvili$^1$ and A.~Rusetsky$^{1,2}$}
\address{
$^1$ High Energy Physics Institute, Tbilisi State University,\\
University St 9, 380086, Tbilisi, Republic of Georgia\\
$^2$ Bogoliubov Laboratory of Theoretical Physics,\\
Joint Institute for Nuclear Research,\\
141980 Dubna (Moscow region), Russia\\
}

\date{\today}

\maketitle

\begin{abstract}\widetext
Bound $q\bar{q}$-systems are studied in the framework of different
3-dimensional relativistic equations derived from the Bethe-Salpeter
equation with the instantaneous kernel in the momentum space.
Except the Salpeter equation, all these equations have a correct one-body
limit when one of the constituent quark masses tends to infinity.
The spin structure of the confining $qq$ interaction potential is taken
in the form $x\gamma_{1}^{0}\gamma_{2}^{0}+(1-x)I_{1}I_{2}$, with
$0\leq x \leq 1$.
At first stage, the one-gluon-exchange potential is
neglected and the confining potential is taken in the oscillator form.
For the systems $(u\bar{s})$, $(c\bar{u})$,
$(c\bar{s})$ and $(u\bar{u})$, $(s\bar{s})$ a comparative qualitative
analysis of these equations is carried out for different values
of the mixing parameter $x$ and the confining potential strength parameter.
We investigate: 1)~the existence/nonexistence of stable solutions of these
equations; 2)~the parameter dependence of the general structure of the
meson mass spectum and leptonic decay constants of pseudoscalar and
vector mesons.
It is demonstrated that none of the 3-dimensional equations considered in
the present paper does
simultaneously describe even general qualitative features of
the whole mass spectrum of $q\bar{q}$ systems. At the same
time, these versions give an acceptable description of the meson
leptonic decay characteristics.
\end{abstract}

\pacs{PACS number(s): 11.10.St, 12.39.Ki, 12.39.Pn}

\section{Introduction}

The Bethe-Salpeter (BS) equation provides a natural basis for the
relativistic treatment of bound $q\bar{q}$ systems in the framework
of the constituent quark model. However, due to the lack of the
probability interpretation of the 4-dimensional (4D) BS amplitude
as well as due to serious mathematical diseases which are inherent
in the BS approach to the bound-state problem, various
3-dimensional (3D) reduction schemes of the original BS equation are
usually used. As it is well known, the simplest version of this sort of
reduction immediately arises if the kernel of the BS equation is taken in
the instantaneous (static) approximation. As a result, the Salpeter
equation is obtained. The Salpeter equation was used for the description
of the bound $q\bar{q}$ system without further approximation in
Refs.~\cite{Long,ACK,FBS,FBS1,Lag,FBSA,Spence,RMMP,4A,Olson,Parramore,PJP},
whereas some additional approximations were
made in Refs.~\cite{Mit,GDD}. However, as it is well known, the
Salpeter equation itself is not free from some drawbacks. Namely, it does not
have a correct one-body limit (the Dirac equation) when the mass of one of
the particles tends to infinity. From the general viewpoint, this property
is expected to be important for the $q\bar{q}$ system with one heavy and one light
quark. In order to avoid the above difficulty, in Refs.~\cite{MW,CJ}
the effective noninteracting 3D Green  function for two fermions was
chosen in the form that guarantees the correct one-body limit of 3D
relativistic equations with the static BS kernel. These versions of the 3D
equations will be referred hereafter as to the MW and CJ versions,
respectively. A new version of the effective free propagator for two scalar
particles which also possesses this property was suggested in Ref.~\cite{MNK}.
The effective 3D Green  function for two noninteracting fermions
can be constructed from this propagator in a standard way.

Taking into account the fact that the relativistic effects are
important for $q\bar{q}$ systems with quarks from light and light-heavy
sectors, it seems interesting to carry out the investigation of this
sort of systems in the framework of the above-mentioned different versions
of 3D relativistic equations. This will allow one to shed light on the
problem of ambiguity
coming from the choice of a particular 3D reduction scheme of the BS
equation, and to find the characteristics of the bound $q\bar q$ systems,
which are more sensitive to this choice.
For the meson mass spectrum this problem was addressed to
in Refs.~\cite{TT1,TT2} where the MW and CJ versions of 3D
relativistic equations together with the Salpeter equation (Sal. version)
were considered in the configuration space to (partially) avoid
the difficulties related to a highly singular behavior of the
linear confining potential in the momentum space at the zero
momentum transfer.
The version of the 3D reduction of the BS equation suggested
in Ref.~\cite{MNK} significantly differs from the MW and CJ
versions and can be written down only in
the momentum space. Consequently, it seems interesting to study together all
versions in the momentum space
and to investigate a wider class of characteristics of the bound systems,
including the decay characteristics of the mesons, which are sensitive to
the behavior of the meson wave functions, and the meson mass spectrum.
These problems will form the subject of the present paper.

The layout of the present paper is as follows. In Section II, we present different
versions of the 3D bound-state equations in the momentum space, and perform
the partial-wave expansion of the obtained equations. The numerical solution
of these equations with the oscillator-type potential is considered in
Section III. The general structure of the meson mass spectra obtained
from the solution of these equations is discussed in detail. In Section
IV, the leptonic decay characteristics of the pseudoscalar and vector
mesons are calculated using the wave functions obtained from the solution
of these equations.

\section{The relativistic 3D equations}

The relativistic 3D equations for the wave function of the bound
$q\bar{q}$ systems, cor\-res\-pon\-ding to the instantaneous
kernel of the BS equation, i.e. when
$K(P;p,p')\rightarrow K_{st}(\vec p,\vec p~')$,
for all versions considered below in the c.m.f. can be written
in a common form
\eq\label{eq1}
        \tilde{\Phi}_ {M}(\vec p\,) =
        \tilde{G}_{0eff}(M,\vec p\,)
        \int\frac{d^3\vec p~'}{(2\pi)^{3}}\,\,
         \bigl[\, iK_{st}(\vec p,\vec p~') \equiv
        \hat{V}(\vec p,\vec p~')\,\bigr]\,\,
        \tilde{\Phi}_{M}(\vec p~')
\en
\noindent
where $M$ is the mass of the bound system, and the equal-time wave function
$\tilde{\Phi}_{M}(\vec p\,)$ is related to the BS amplitude $\Phi_{P}(p)$ as
\eq
       \tilde{\Phi}_{M}(\vec p\,) =
       \int\frac{dp_{0}}{2\pi}\,\Phi_{P=(M,\vec 0\,)}(p)
\en
\noindent
The effective 3D Green  function of two noninteracting-quark system
$\tilde{G}_{0eff}$ is defined as
\eq\label{eq3}
\tilde{G}_{0eff}(M,\vec p\,)=\int\frac{dp_0}{2\pi i}\,
\bigl[\, G_{0eff}(M;p) =
g_{0eff}(M;p)(\not\! p_{1}+m_{1})(\not\! p_{2}+m_{2})\,\bigr]
\en
\noindent
Here
$g_{0eff}(M;p)$ is the effective propagator of two scalar particles. The
operator $\tilde{G}_{0eff}$ is given in the form
\eq\label{eq4}
\tilde{G}_{0eff}(M,\vec p\,) =
\sum\limits_{\alpha_{1}=\pm}\sum\limits_{\alpha_{2}=\pm}\,
\frac{D^{(\alpha_{1}\alpha_{2})}(M;p)}{d(M;p)}\,\,
\Lambda_{12}^ {(\alpha_1\alpha_2)}
(\vec p,-\vec p\,)\,\,\gamma_{1}^{0}\gamma_{2}^{0},\,\,\,\,\,\,\,\,
p\equiv|\vec p\,|
\en
\noindent
where the projection operators $\Lambda_{12}^{(\alpha_1\alpha_2)}$
are defined by
\eqa
 \Lambda_{12}^{(\alpha_1\alpha_2)}(\vec p_{1},\vec p_{2})
=\Lambda_{1}^{(\alpha_1)}(\vec p_{1})\otimes
 \Lambda_{2}^{(\alpha_2)}(\vec p_{2})&,&\,\,\,\,\,
 \Lambda_{i}^{(\alpha_i)}(\vec p_{i})
=\frac{\omega_{i}+\alpha_i\hat h_{i}(\vec p_{i})}{2\omega_{i}}
\nonumber\\[2mm]
\hat h_{i}(\vec p_{i})=\gamma_{i}^{0}(\vec\gamma_{i}\,\vec p_{i})+
m_{i}\gamma_{i}^{0}&,&\,\,\,\,\,\,\,\,
\omega_{i}= \bigl( m_{i}^{2}+\vec p_{i}^{~2}\bigr)^{1/2}
\ena
\noindent
and the functions $D^{(\alpha_{1}\alpha_{2})}(M;p)$ and $d(M;p)$ are
given by the expressions (see Ref.~\cite{BKA})
\eqa
D^{(\alpha_{1}\alpha_{2})}=
\frac{(-1)^{\alpha_{1}+\alpha_{2}}}
{\omega_{1}+\omega_{2}-(\alpha_{1}E_{1}+\alpha_{2}E_{2})},
\,\,\,\,\,
d=1
&\nonumber\\[2mm]
E_{1}+E_{2}=M,\,\,\,\,
E_{1}-E_{2}=\frac{m_{1}^{2}-m_{2}^{2}}{M}\equiv b_0
&{\mbox{\quad(MW version)}}\\[2mm]
D^{(\alpha_{1}\alpha_{2})}=
(E_{1}+\alpha_{1}\omega_{1})(E_{2}+\alpha_{2}\omega_{2})
&\nonumber\\[2mm]
d=2(\omega_{1}+\omega_{2})a,\,\,\,\,a=
E_{i}^{2}-\omega_{i}^{2}=[M^{2}+b_{0}^{2}-2(\omega_{1}^{2}+\omega_{2}^{2})]/4
&\mbox{\quad(CJ version)}
\label{eq7}
\ena
\noindent

Note that for the case of CJ version Eq. (\ref{eq7}) is obtained from
Eqs. (\ref{eq3}) and (\ref{eq4})
by using the expression for $g_{0eff}(M;p)$  defined
from the dispersion relation which guarantees the elastic unitarity.
The same condition is satisfied by the expression of
$g_{0eff}(M;p)$ suggested in Ref.~\cite{MNK}, (see formula (10) from
this paper). According to this condition,
the particles in the intermediate states are allowed to go off shell
proportionally to their mass, so that when one of the particles becomes
infinitely massive, it automatically keeps that  particle fully on-mass-shell
and the equation is reduced to the one-body equation. Using this
expression for $g_{0eff}(M;p)$ in Eq. (\ref{eq3}), we derive the expression
for $\tilde{G}_{0eff}$ having the form (\ref{eq4}) where
\eqa
&&
D^{(\alpha_{1}\alpha_{2})}=
(E_{1}+\alpha_{1}\omega_{1})(E_{2}+\alpha_{2}\omega_{2})-\frac{R-b}{2y}
\biggl[\frac{R-b}{2y}
+(E_{1}+\alpha_{1}\omega_{1})-(E_{2}+\alpha_{2}\omega_{2})\biggr]
\nonumber\\[2mm]
&&
d=2RB,
\hspace*{0.5cm}
R=\bigl(b^{2}-4y^{2}a\bigr)^{1/2},
\hspace*{0.5cm}
B=\frac{R-b}{2y}\biggl[\frac{R-b}{2y}+b_{0}\biggr]+a,
\nonumber\\[2mm]
&&
b=M+b_{0}y,
\hspace*{0.5cm}
y=\frac{m_{1}-m_{2}}{m_{1}+m_{2}}
\label{eq8}
\ena
Hereafter this version will be referred to as the MNK version
(our comments on Ref.~\cite{MNK} see in Ref.~\cite{BKA}).

Using the properties of the projection operators
$\Lambda_{12}^ {(\alpha_1\alpha_2)}$ and Eqs.~(\ref{eq4})-(\ref{eq8}),
the following system of equations can be derived from Eq. (\ref{eq1})
\eqa
&&[M-(\alpha_{1}\omega_{1}+\alpha_{2}\omega_{2})]
\tilde{\Phi}_ {M}^{(\alpha_{1}\alpha_{2})}(\vec p\,)=
\nonumber\\[2mm]
&=&A^{(\alpha_{1}\alpha_{2})}(M;p)\,
\Lambda_{12}^{(\alpha_1\alpha_2)}(\vec p,-\vec p\,)
\int\frac{d^3\vec p~'}{(2\pi)^{3}}\,
\gamma_{1}^{0}\gamma_{2}^{0}\,
\hat{V}(\vec p,\vec p~')
\sum\limits_{\alpha_{1}^{'}=\pm}\,\sum\limits_{\alpha_{2}^{'}=\pm}\,
\tilde{\Phi}_ {M}^{(\alpha_{1}^{'}\alpha_{2}^{'})}(\vec p~')
\label{eq9}
\ena
\noindent
where $\tilde{\Phi}_ {M}^{(\alpha_{1}\alpha_{2})}(\vec p\,)=
\Lambda_{12}^{(\alpha_1\alpha_2)}(\vec p,-\vec p\,)\
\tilde{\Phi}_ {M}(\vec p)$
and the functions $A^{(\alpha_{1}\alpha_{2})}(M;p)$ are given by
\eqa
A^{(\pm\pm)}=\pm 1,
\hspace*{1.cm}
A^{(\pm\mp)}=\frac{M}{\omega_{1}+\omega_{2}}
&&{\mbox{\quad(MW version)}}
\label{eq10}
\\[2mm]
A^{(\alpha_{1}\alpha_{2})}=
\frac{M+(\alpha_{1}\omega_{1}+\alpha_{2}\omega_{2})}
{2(\omega_{1}+\omega_{2})}
&&{\mbox{\quad(CJ version)}}
\label{eq11}
\\[2mm]
A^{(\alpha_{1}\alpha_{2})}=\frac{1}{2RB}\biggl\{
a[M+(\alpha_{1}\omega_{1}+\alpha_{2}\omega_{2})] - \hspace*{4.2cm} &
\nonumber\\[2mm]
-[M-(\alpha_{1}\omega_{1}+\alpha_{2}\omega_{2})]
\frac{R-b}{2y}
\biggl[\frac{R-b}{2y}+(E_{1}+\alpha_{1}\omega_{1})-
(E_{2}+\alpha_{2}\omega_{2})\biggr]\biggr\}
&&{\mbox{\quad(MNK version)}}
\label{eq12}
\ena
As to the Salpeter equation, it can be obtained from the MW version
by putting $A^{(\pm\mp)}=0$ and $\tilde{\Phi}_{M}^{(\pm\mp)}=0$.

Note that for quarks with the equal masses $(m_{1}=m_{2}=m)$
and $\omega=\bigl( m^{2}+\vec p^{~2}\bigr)^{1/2}$, from
Eqs. (\ref{eq10}) and (\ref{eq11}) we have
\eqa
A^{(\pm\pm)}=\pm 1,
\hspace*{1.cm}
A^{(\pm\mp)}=\frac{M}{2\omega}
\hspace*{1.cm}
&&{\mbox{(MW version)}}
\label{eq13}
\\[2mm]
A^{(\pm\pm)}=
\frac{M+2\omega}{4\omega}
\hspace*{1.cm}
A^{(\pm\mp)}=\frac{M}{4\omega}
\hspace*{1.cm}
&&{\mbox{(CJ version)}}
\label{eq14}
\ena
One observes from Eqs. (\ref{eq13}) and (\ref{eq14}) that in the limit of
the equal-mass quarks the bound-state mass $M$ enters multiplicatively in the
coefficients in front of the mixed-energy components
$\tilde{\Phi}_{M}^{(\pm\mp)}(\vec p\,)$ both in the l.h.s. and r.h.s.
of Eq. (\ref{eq9}). Consequently, dividing both sides of the
equations for these components by $M$, one arrives at the (nondynamical)
constraints on all components of the wave function
which must be considered together with the remaining two dynamical equations
for the components $\tilde{\Phi}_{M}^{(\pm\pm)}(\vec p\,)$.
These equations for the bound state mass $M$
are linear in the MW version and are nonlinear in the CJ version due to the fact
that the r.h.s. of the equations depends on the value of $M$ one is looking
for. In the MNK version with an account of the property of the function
$R$ given by Eq. (\ref{eq8})
$\lim\limits_{m_{1}\rightarrow m_{2}}
R/y=\lim\limits_{m_{1}\rightarrow m_{2}}M/y$,
from Eq. (\ref{eq12}) we have
\begin{eqnarray}
A^{(\pm\pm)}=
\frac{M+2\omega}{2\omega}
\hspace*{1.2cm}
A^{(\pm\mp)}=\frac{1}{2}
&&{\mbox{\quad\quad(MNK version)}}
\end{eqnarray}

In this case, from Eq. (\ref{eq9}) it follows that characteristic features
of the bound-state equations remain
unchanged -- again it is the system of nonlinear equations in $M$ and
it includes all components of the wave function
$\tilde{\Phi}_{M}^{(\alpha_{1}\alpha_{2})}(\vec p\,)$, as it is in the case
for the quarks with nonequal masses.

Further, we write the unknown function
$\tilde{\Phi}_{M}^{(\alpha_{1}^{'}\alpha_{2}^{'})}$
in Eq. (\ref{eq9}) in the form analogous to that used in Ref.~\cite{4A},
where the bound $q\bar{q}$ systems were studied in the framework of the
Salpeter equation
\eqa
&&
\tilde{\Phi}_ {M}^{(\alpha_{1}\alpha_{2})}(\vec p\,)=
N_ {12}^{(\alpha_{1}\alpha_{2})}(p)
\pmatrix{1\cr
 \alpha_{1}(\vec \sigma_{1}\vec p\,)/(\omega_{1}+\alpha_{1}m_{1})
}
\otimes
\pmatrix{1\cr
-\alpha_{2}(\vec \sigma_{2}\vec p\,)/(\omega_{2}+\alpha_{2}m_{2})
}
\chi_{M}^{(\alpha_{1}\alpha_{2})}(\vec p\,)
\nonumber\\
&&\label{eq16}
\ena
\noindent where
\eq\label{eq17}
N_ {12}^{(\alpha_{1}\alpha_{2})}(p)=
\biggl(\frac{\omega_{1}+\alpha_{1}m_{1}}{2\omega_{1}}\biggr)^{1/2}
\biggl(\frac{\omega_{2}+\alpha_{2}m_{2}}{2\omega_{2}}\biggr)^{1/2}
\equiv N_ {1}^{(\alpha_{1})}(p)N_ {2}^{(\alpha_{2})}(p)
\en
Then, if the $qq$ interaction potential operator
$\hat{V}(\vec p,\vec p~')$ is taken in the form \cite{4A}
\eq\label{eq18}
\hat{V}(x;\vec p,\vec p~')=\gamma_{1}^{0}\gamma_{2}^{0}
\hat{V}_{og}(\vec p - \vec p~')+
[x\gamma_{1}^{0}\gamma_{2}^{0}+(1-x)I_{1}I_{2}]
\hat{V}_{c}(\vec p - \vec p~'),
\hspace*{0.8cm}
(0\leq x \leq1),
\en
\noindent
the following system of equations for the Pauli $2\otimes2$ wave
functions $\chi_{M}^{(\alpha_{1}\alpha_{2})}$  can be derived
\eqa
&&
[M-(\alpha_{1}\omega_{1} + \alpha_{2}\omega_{2})]
\chi_{M}^{(\alpha_{1}\alpha_{2})}(\vec p\,)=
\nonumber\\[2mm]
&=&A^{(\alpha_{1}\alpha_{2})}(M;p)
\sum\limits_{\alpha_{1}^{'}=\pm}\,\sum\limits_{\alpha_{2}^{'}=\pm}\,
\int\frac{d^3\vec p~'}{(2\pi)^{3}}\,
\hat{V}_{eff}^{(\alpha_{1}\alpha_{2},\alpha_{1}^{'}\alpha_{2}^{'})}
(\vec p,\vec p~',\vec \sigma_{1},\vec \sigma_{2})
\chi_{M}^{(\alpha_{1}^{'}\alpha_{2}^{'})}(\vec p~')
\label{eq19}
\ena
\noindent
where
\eqa
\hat{V}_{eff}^{(\alpha_{1}\alpha_{2},\alpha_{1}^{'}\alpha_{2}^{'})}
(\vec p,\vec p~',\vec \sigma_{1},\vec \sigma_{2})
&=&
N_ {12}^{(\alpha_{1}\alpha_{2})}(p)\biggl[V(1;\vec p-\vec p~')
\hat{B}_{1}^{(\alpha_{1}\alpha_{2},\alpha_{1}^{'}\alpha_{2}^{'})}
(\vec p,\vec p~',\vec \sigma_{1},\vec \sigma_{2})
\nonumber\\[2mm]
&+&V(x;\vec p-\vec p~')
\hat{B}_{2}^{(\alpha_{1}\alpha_{2},\alpha_{1}^{'}\alpha_{2}^{'})}
(\vec p,\vec p~',\vec \sigma_{1},\vec \sigma_{2})
\biggr]
N_ {12}^{(\alpha_{1}^{'}\alpha_{2}^{'})}(p')
\ena
\eq
\hat{B}_{1}^{(\alpha_{1}\alpha_{2}\alpha_{1}^{'}\alpha_{2}^{'})}=1+
\frac{\alpha_{1}\alpha_{2}\alpha_{1}^{'}\alpha_{2}^{'}
({\vec{\sigma}}_{1}{\vec p}\,)({\vec{\sigma}}_{2}{\vec p}\,)
({\vec{\sigma}}_{1}{\vec p~'})({\vec{\sigma}}_{2}{\vec p~'})}
{(\omega_{1}+\alpha_{1} m_{1})(\omega_{2}+\alpha_{2} m_{2})
(\omega_{1}^{'}+\alpha_{1}^{'} m_{1})(\omega_{2}^{'}+\alpha_{2}^{'} m_{2})}
\en
\eq
\hat{B}_{2}^{(\alpha_{1}\alpha_{2}\alpha_{1}^{'}\alpha_{2}^{'})}=
\frac{\alpha_{1}\alpha_{1}^{'} {(\vec{\sigma}}_{1}{\vec p}\,)
({\vec{\sigma}}_{1}{\vec p~'})}{(\omega_{1}+\alpha_{1} m_{1})
(\omega_{1}^{'}+\alpha_{1}^{'} m_{1})}+
\frac{\alpha_{2}\alpha_{2}^{'} ({\vec{\sigma}}_{2}{\vec p}\,)
({\vec{\sigma}}_{2}{\vec p~'})}{(\omega_{2}+\alpha_{2} m_{2})
(\omega_{2}^{'}+\alpha_{2}^{'} m_{2})}
\en
\eq
V(x;\vec p-\vec p~')=V_{og}(\vec p-\vec p~')
+(2x-1)V_{c}(\vec p-\vec p~')
\en
\noindent
Now using the partial-wave expansion
\eq\label{eq24}
\chi_{M}^{(\alpha_{1}\alpha_{2})} (\vec p\,)=
\sum\limits_{LSJM_{J}}\bigl<\hat{p}\mid LSJM_{J} \bigr>
R_{LSJ}^{(\alpha_{1}\alpha_{2})}(p)
,\,\,\,\,\,\,\,\,\biggl(\hat{p}=\frac{\vec p}{p}\biggr)
\en
\noindent
\eq
V(\vec p-\vec p~')=(2\pi)^{3}\sum\limits_{LSJM_{J}}V^{L}(p,p')
\bigl<\hat{p}\mid LSJM_{J} \bigr>\bigl<LSJM_{J}\mid\hat{p}\bigr>
\en
\noindent
where
\eq\label{eq26}
V^{L}(p,p^{'})=\frac{2}{\pi}\int_{0}^{\infty}j_{L}(pr)j_{L}(p^{'}r)r^{2}dr
\en
\noindent
with $j_{L}(x)$ being the spherical Bessel function,
the following system of equations can
be obtained from Eq. (\ref{eq19})
for the radial wave functions
$R_{LSJ}^{(\alpha_{1}\alpha_{2})}(p)$
\eqa
&&[M-(\alpha_{1}\omega_{1}+\alpha_{2}\omega_{2})]
R_{J(^{0}_{1})J}^{(\alpha_{1}\alpha_{2})}(p)=
A^{(\alpha_{1}\alpha_{2})}(M;p)N_ {12}^{(\alpha_{1}\alpha_{2})}(p)
\times
\nonumber\\[2mm]
&\times&
\sum\limits_{\alpha_{1}^{'}\alpha_{2}^{'}}
\int_{0}^{\infty} p^{'2}dp^{'}\biggl \{\biggl[\biggl(1+a_{12 \otimes 12}^
{(\alpha_{1}\alpha_{2},\alpha_{1}^{'}\alpha_{2}^{'})}(p,p{'})\biggr)
V^{J}(1;p,p^{'})+
\nonumber\\[2mm]
&+&a_{\oplus}^{(\alpha_{1}\alpha_{2},\alpha_{1}^{'}
\alpha_{2}^{'})}(p,p^{'})V^{(^{0}_{1})}_{\oplus J}(x;p,p^{'})\biggr]
R_{J(^{0}_{1})J}^{(\alpha_{1}^{'}\alpha_{2}^{'})}(p^{'})-
\nonumber\\[2mm]
&-&\biggl[a_{\ominus}^{(\alpha_{1}\alpha_{2},\alpha_{1}^{'}\alpha_{2}^{'})}(p,p^{'})
V_{\ominus J}(x;p,p^{'})\biggr]
R_{J(^{1}_{0})J}^{(\alpha_{1}^{'}\alpha_{2}^{'})}(p^{'})\biggr\}
N_ {12}^{(\alpha_{1}^{'}\alpha_{2}^{'})}(p')
\nonumber
\ena

\eqa
&&
[M-(\alpha_{1}\omega_{1}+\alpha_{2}\omega_{2})]
R_{J\pm11J}^{(\alpha_{1}\alpha_{2})}(p)=
A^{(\alpha_{1}\alpha_{2})}(M;p)N_ {12}^{(\alpha_{1}\alpha_{2})}(p)
\times
\nonumber\\[2mm]
&\times&
\sum\limits_{\alpha_{1}^{'}\alpha_{2}^{'}}
\int_{0}^{\infty} p^{'2}dp^{'}\biggl\{\biggl[V^{J\pm1}(1;p,p^{'})
+a_{12 \otimes 12}^{(\alpha_{1}\alpha_{2},
\alpha_{1}^{'}\alpha_{2}^{'})}(p,p{'})V_{J\pm1 1 J}(1;p,p^{'})+
\nonumber\\[2mm]
&+&a_{\oplus}^{(\alpha_{1}\alpha_{2},\alpha_{1}^{'}
\alpha_{2}^{'})}(p,p^{'})V^{J}(x;p,p^{'})\biggr]
R_{J\pm11J}^{(\alpha_{1}^{'}\alpha_{2}^{'})}(p^{'})+
\nonumber\\[2mm]
&+&\biggl[a_{12 \otimes 12}^{(\alpha_{1}\alpha_{2},\alpha_{1}^{'}\alpha_{2}^{'})}
(p,p^{'})\frac{2}{2J+1}V_{\ominus J}(1;p,p^{'})\biggr]
R_{J\mp11J}^{(\alpha_{1}^{'}\alpha_{2}^{'})}(p^{'})\biggr\}
N_ {12}^{(\alpha_{1}^{'}\alpha_{2}^{'})}(p')
\label{eq27}
\ena
\noindent where
\eqa
a_{12 \otimes 12}^{(\alpha_{1}\alpha_{2},
\alpha_{1}^{'}\alpha_{2}^{'})}(p,p{'})
&=&a_{12}^{(\alpha_{1}\alpha_{2})}(p,p)
a_{12}^{(\alpha_{1}^{'}\alpha_{2}^{'})}(p^{'},p^{'})
\nonumber\\[2mm]
a_{^{\oplus}_{\ominus}}^{(\alpha_{1}\alpha_{2},\alpha_{1}^{'}\alpha_{2}^{'})}
(p,p{'})&=&a_{11}^{(\alpha_{1}\alpha_{1}^{'})}(p,p^{'})\pm
a_{22}^{(\alpha_{2}\alpha_{2}^{'})}(p,p^{'})
\nonumber\\[2mm]
a_{ij}^{(\alpha_{i}\alpha_{j})}(p,p^{'})=a_{i}^{\alpha_{i}}(p)
a_{j}^{\alpha_{j}}(p^{'})&,&
\hspace*{1.2cm}
a_{i}^{\alpha_{i}}(p)=\frac{\alpha_{i}p}{\omega_{i}+\alpha_{i}m_{i}}
\ena
and
\eqa
V^{(^{0}_{1})}_{\oplus J}(x;p,p')&=&\frac{1}{2J+1}\bigl[
\pmatrix{J\cr J+1}V^{J-1}(x;p,p')+
\pmatrix{J+1\cr J}V^{J+1}(x;p,p')
\bigr]
\nonumber\\[2mm]
V_{\ominus J}(x;p,p')&=&\frac{\sqrt{J(J+1)}}{2J+1}\bigl[
V^{J-1}(x;p,p')-V^{J+1}(x;p,p')\bigr]
\\[2mm]
V_{J\pm1 1 J}(1;p,p')&=&\frac{1}{(2J+1)^{2}}\bigl[
V^{J\pm1}(1;p,p')+4J(J+1)V^{J\mp1}(1;p,p')\bigr]
\nonumber\\[2mm]
V(x;p,p')&=&V_{og}(x;p,p')+(2x-1)V_{c}(x;p,p')
\nonumber
\ena

The main purpose of the present study is to carry out the comparative
qualitative ana\-ly\-sis of the different versions of 3D relativistic equations
(\ref{eq27}), addressing the question of existence of stable solutions
for different values of the scalar-vector
mixing parameter $x$ in the confining part of the potential (Eq. (\ref{eq18})).
Also, we investigate the general structure of the meson mass spectrum
and calculate the leptonic decay characteristics,
namely, the pseudoscalar decay constant $f_{P}(P \rightarrow\mu \bar{\nu})$
and the vector meson decay width $\Gamma (V \rightarrow e^{-} e^{+})$.
For these reasons, at the first stage we neglect in (\ref{eq18}) the one-gluon
exchange potential. A full analysis of the problem will
be made in further publications.
Further, according to Ref.~\cite{2A}, we use the oscillator
form for the confining potential $V_{c}(r)$
which is a simplified, but justified
form of a more general potential used in Ref.~\cite{4A}
(at least for the quarks from the light and light-heavy sectors,
which are considered in the present paper). Namely, we take
\eqa
V_{c}(r)=\frac{4}{3}\alpha_{s}(m_{12}^{2})\biggl(\frac{\mu_{12}\omega_{0}^{2}}
{2}r^{2}-V_{0}\biggr)
\\[2mm]
\mu_{12}=\frac{m_{1}m_{2}}{m_{12}},
\hspace*{1.cm}
m_{12}=m_{1}+m_{2},
\hspace*{1.cm}
\alpha_{s}(Q^{2})=\frac{12\pi}{33-2n_f}\biggl({\mbox{\rm ln}}\frac{Q^{2}}{\Lambda^{2}}
\biggr)^{-1}
\nonumber
\ena
where $n_f$ denotes the number of flavors. This potential
in the momentum space corresponds to the operator (see (\ref{eq26}))
\eq\label{eq30}
V_{c}^{L}(p,p^{'})=-\frac{4}{3}\alpha_{s}(m_{12}^{2})
\biggl[\frac{\mu_{12}\omega_{0}^{2}}{2}\bigl(\frac{d^{2}}{dp^{'2}}+
\frac{2}{p^{'}}\frac{d}{dp^{'}}-\frac{L(L+1)}{p^{'2}}\bigr)+V_{0}\biggr]
\frac{\delta(p-p^{'})}{pp'}
\en
Now the system of equations (\ref{eq27}) can be reduced to the
system of equations with the following structure:

\eqa
&&
[M-(\alpha_{1}\omega_{1}+\alpha_{2}\omega_{2})]
R_{J(^{0}_{1})J}^{(\alpha_{1}\alpha_{2})}(p)=
-\frac{4}{3}\alpha_{s}(m_{12}^{2}) A^{(\alpha_{1}\alpha_{2})}(M;p)
\sum\limits_{\alpha_{1}^{'}\alpha_{2}^{'}} \biggl\{ V_{0}
\biggl[B_{\oplus}^{(\alpha_{1}\alpha_{2},\alpha_{1}^{'}\alpha_{2}^{'})}(p)+
\nonumber\\[2mm]
&&
+(2x-1)A_{\oplus}^{(\alpha_{1}\alpha_{2},\alpha_{1}^{'}\alpha_{2}^{'})}(p)
\biggr]R_{J(^{0}_{1})J}^{(\alpha_{1}^{'}\alpha_{2}^{'})}(p^{'})+
\frac{\mu_{12}\omega_{0}^{2}}{2}\biggl[
\biggl(\hat{D}_{B}^{(\alpha_{1}\alpha_{2},\alpha_{1}^{'}\alpha_{2}^{'})}(p)+
(2x-1)\hat{D}_{A}^{(\alpha_{1}\alpha_{2},\alpha_{1}^{'}\alpha_{2}^{'})}(p)-
\nonumber\\[2mm]
&&
-\frac{1}{p^{2}}
\biggl(B_{\oplus}^{(\alpha_{1}\alpha_{2},\alpha_{1}^{'}\alpha_{2}^{'})}(p)J(J+1)
+(2x-1)A_{\oplus}^{(\alpha_{1}\alpha_{2},\alpha_{1}^{'}\alpha_{2}^{'})}(p)
\pmatrix{J(J+1)+2\cr J(J+1)}
\biggr)\biggr)
R_{J(^{0}_{1})J}^{(\alpha_{1}^{'}\alpha_{2}^{'})}(p^{'})-
\nonumber\\[2mm]
&&
-(2x-1)\biggl(\frac{2\sqrt{J(J+1)}}{p^{2}}
A_{\ominus}^{(\alpha_{1}\alpha_{2},\alpha_{1}^{'}\alpha_{2}^{'})}(p)\biggr)
R_{J(^{1}_{0})J}^{(\alpha_{1}^{'}\alpha_{2}^{'})}(p)\biggr]\biggr\},
\nonumber\\[2mm]
&&
[M-(\alpha_{1}\omega_{1}+\alpha_{2}\omega_{2})]
R_{J\pm11J}^{(\alpha_{1}\alpha_{2})}(p)=
-\frac{4}{3}\alpha_{s}(m_{12}^{2}) A^{(\alpha_{1}\alpha_{2})}(M;p)
\sum\limits_{\alpha_{1}^{'}\alpha_{2}^{'}}\biggl\{V_{0}
\biggl[B_{\oplus}^{(\alpha_{1}\alpha_{2},\alpha_{1}^{'}\alpha_{2}^{'})}(p)+
\nonumber\\[2mm]
&&
+(2x-1)A_{\oplus}^{(\alpha_{1}\alpha_{2},\alpha_{1}^{'}\alpha_{2}^{'})}(p)
\biggr] R_{J\pm11J}^{(\alpha_{1}\alpha_{2})}(p)
+\frac{\mu_{12}\omega_{0}^{2}}{2}\biggl[
\bigl(\hat{D}_{B}^{(\alpha_{1}\alpha_{2},\alpha_{1}^{'}\alpha_{2}^{'})}(p)+
(2x-1)\hat{D}_{A}^{(\alpha_{1}\alpha_{2},\alpha_{1}^{'}\alpha_{2}^{'})}(p)-
\nonumber\\[2mm]
&&
-\frac{1}{p^{2}}
\biggl(B_{\oplus}^{(\alpha_{1}\alpha_{2},\alpha_{1}^{'}\alpha_{2}^{'})}(p)
\biggl(J(J+1)+1\pm\frac{1}{2J+1}\biggr)\pm\frac{4J(J+1)}{2J+1}
B_{\ominus}^{(\alpha_{1}\alpha_{2},\alpha_{1}^{'}\alpha_{2}^{'})}(p)+
\nonumber\\[2mm]
&&
+(2x-1)A_{\oplus}^{(\alpha_{1}\alpha_{2},\alpha_{1}^{'}\alpha_{2}^{'})}(p)
J(J+1)\biggr)\biggr)R_{J\pm11J}^{(\alpha_{1}\alpha_{2})}(p)+
\nonumber\\[2mm]
&&
+\frac{1}{p^{2}}\frac{2\sqrt{J(J+1)}}{(2J+1)^{2}}\biggl(
B_{\oplus}^{(\alpha_{1}\alpha_{2},\alpha_{1}^{'}\alpha_{2}^{'})}(p)-
B_{\ominus}^{(\alpha_{1}\alpha_{2},\alpha_{1}^{'}\alpha_{2}^{'})}(p)\biggr)
R_{J\mp11J}^{(\alpha_{1}\alpha_{2})}(p)\biggr]\biggr\}
\label{eq31}
\ena
\noindent
where $A_{^{\oplus}_{\ominus}}^{(\alpha_{1}\alpha_{2},\alpha_{1}^,\alpha_{2}^{'})} and
B_{^{\oplus}_{\ominus}}^{(\alpha_{1}\alpha_{2},\alpha_{1}^{'}\alpha_{2}^{'})}$
are a given functions of $p$, and
$\hat{D}_{A(B)}^{(\alpha_{1}\alpha_{2},\alpha_{1}^{'}\alpha_{2}^{'})}$
are certain second order differential operators in $p$.

\section{Meson mass spectrum}

In order to solve the system of equations (\ref{eq31}) for the bound state mass $M$,
the unknown radial wave functions $R_{LSJ}^{(\alpha_{1}\alpha_{2})}(p)$ are
expanded in the basis of the radial wave functions $R_{nL}(p)$ being the solutions
of the Shr\"{o}dinger radial equation with the oscillator potential in the momentum space
(\ref{eq30}), as it was done in Refs.~\cite{4A,2A}
\eq\label{eq32}
R_{LSJ}^{(\alpha_{1}\alpha_{2})}(p)=\bigl( 2M(2\pi)^{3}\bigr)^{1/2}
\sum\limits_{n=0}^{\infty}C_{LSJn}^{(\alpha_{1}\alpha_{2})}R_{nL}(p)
\en
where the multiplier  $\bigl( 2M(2\pi)^{3}\bigr)^{1/2}$ is introduced for appropriate
normalization of the wave function and
\eqa
&&
R_{nL}(p)=p_{0}^{-\frac{3}{2}}\biggl[\biggl(\frac{2 \Gamma (n+1+\frac{3}
{2})}{\Gamma (n+1)}\biggr )^{1/2}\frac{z^{L}\exp\bigl(- \frac{z^{2}}{2}\bigr)}
{\Gamma(L+\frac{3}{2})}\,\,_{1}F_{1}(-n,L+\frac{3}{2};z^{2})
\equiv \bar R_{nL}(z)\biggr]
\nonumber\\[2mm]
&&
z=\frac{p}{p_{0}},
\hspace*{1.cm}
p_{0}=\biggl(\mu_{12}\omega_{0}
\biggl(\frac{4}{3}\alpha_{s}(m_{12}^{2})\biggr)^{1/2}\biggr)^{1/2}
\label{eq33}
\ena
Substituting the expression (\ref{eq32}) into the system of differential equations
(\ref{eq31}), the following system of algebraic equations for the coefficients
$C_{LSJn}^{\alpha_{1}\alpha_{2}}$ can be obtained
\eq\label{eq34}
MC_{LSJn}^{(\alpha_{1}\alpha_{2})}=
\sum\limits_{\alpha_{1}^{'}\alpha_{2}^{'}}\sum\limits_{L^{'}S^{'}n^{'}}
H^{(\alpha_{1}\alpha_{2},\alpha_{1}^{'}\alpha_{2}^{'})}_{LSJn,
L^{'}S^{'}n^{'};J}(M)C_{L^{'}S^{'}Jn^{'}}^{(\alpha_{1}^{'}\alpha_{2}^{'})}
\en

It is necessary to note here that the matrix $H_{\alpha\beta}$ explicitly
depends (except the Sal. version) on the meson mass $M$ we are looking for.
Consequently, the system of equations (\ref{eq34}) is nonlinear in $M$. Note
also that for the quarks with equal masses $(m_{1}=m_{2}=m)$ part of the
equations from the system (\ref{eq34}), corresponding to $(\alpha_{1}\alpha_{2})=(\pm \mp)$
for the MW and CJ versions, transforms into the nondynamical constraints between
all coefficients
$C_{LSJn}^{(\alpha_{1}\alpha_{2})}$ which should be considered together with
the remaining dynamical equations corresponding to
$(\alpha_{1}\alpha_{2})=(\pm\pm)$.

The numerical algorithm for the solution of the system of nonlinear
equations (\ref{eq34}) in the case of
nonequal mass quarks was discussed in Ref.~\cite{BKA} where the systems
$u\bar{s}$ $(^{3}S_{1}, ^{1}P_{1}, ^{3}P_{0},^{3}P_{1},
^{3}P_{2},^{1}D_{2}, ^{3}D_{1}, ^{3}D_{3})$, $c\bar{u}$ and $c\bar{s}$
$(^{1}S_{0}, ^{3}S_{1}, ^{1}P_{1}, ^{3}P_{2})$ were considered.
In brief, the infinite set of equations (\ref{eq34}) is truncated at
some fixed value $n=N_{max}$ and the eigenvalue problem is solved for
the finite-range matrix $H$. Increasing then $N_{max}$, one checks the
stability of the resulting spectrum with respect to the variation of
$N_{max}$. Since the r.h.s. of Eq. (\ref{eq34}) depends on $M$, the
solutions are obtained iteratively, starting from some value of $M$.
In Ref. \cite{BKA}, the existence of
stable solutions of the system of equations (\ref{eq34}) was investigated
for different values of the mixing parameter
$x$ in the confining potential (\ref{eq18}). It was found that
the existence/nonexistence of stable solutions of Eq. (\ref{eq34})
critically depended on the value of $x$, on the value of confining interaction
strength parameter $\omega_{0}$ (\ref{eq30}), and on the particular state
under consideration.
This dependence is different for the different versions of 3D reduction of the BS
equation. The instability is primarily caused by the presence of the mixed
$(+-,-+)$ energy
components of the wave function in the equations for the $q\bar{q}$ bound
system. Namely, for the parameter $\omega_{0}$=710 MeV that leads
to a reasonable description of the meson mass spectrum in the framework
of the Salpeter equation \cite{4A}, stable solutions for CJ, MNK and Sal.
versions simultaneously exist for the values of the parameter $x$
from the interval $0.3\leq x \leq 0.6-0.7$. For the MW version,
in order to provide the existence of stable solutions in the
same interval of $x$, $\omega_{0}$ must be set to a smaller value
(450 MeV). However, in this case the values of masses for all states under
consideration turn out to be smaller than the experimental ones.
Keeping in mind that
the calculated values of masses will further decrease
after adding the one-gluon interaction potential,
we conclude that the MW version seems to poorly describe the meson mass
spectrum. For this reason, along with the Sal. version, below we consider
only the CJ and MNK versions, both having a similar theoretical foundation:
the effective Green  function (\ref{eq3}) in these versions is constructed
from the elastic unitarity condition.

The results of calculations are given in Figs. 1,2,3, from which one can see
that the level ordering is similar for all three versions under
consideration. Further, at $x=0.5$ the states $^{3}P_{0},^{3}P_{1},
^{1}P_{1}$ are degenerate and spin-orbit splitting exists only between the
degenerate $^{3}P_{0},^{3}P_{1}$ states and the $^{3}P_{2}$ state. For
$x\neq0.5$ this degeneration is removed and the calculated level ordering
agrees with the experiment for the
value $x=0.3$, except the $^{3}P_{2}$ state. For the D - states
$(u\bar{s})$, experimentally there is degeneration between $^{1}D_{2},^{3}D_{3}$
states. In our calculations we do not have this degeneration,
but for the MNK and CJ
versions at $x=0.3$ the splitting is very small and increases for $x=0.5$
and $x=0.7$. Note, however, that only for these values of $x$ the sequence
of the $^{3}D_{1}$ state and other two
D-states agrees with the experiment. For the $q\bar{q}$ states with the quarks
from light-heavy sectors $(c\bar{u},c\bar{s})$ the spin-dependence of the energy
levels for all values of $x$ is much weaker than the experimental one,
but at the same time for $x \geq 0.5$ the level ordering agrees with
experimental data.

As to the pseudoscalar $q\bar{q}$ systems (the $^{1}S_{0}$ state) with the quarks from
the light sector ($u\bar{s}$), the calculated masses in the model under
consideration are much larger than the experimental ones, as demonstrated in
Fig 1. This might serve as an indication of the fact that if the constituent
quark model is used for the description of this sort of systems,
the chiral symmetry breaking effect should be taken into
account, at least in a phenomenological manner,
e.g. by the inclusion of the t'Hooft interaction in the kernel
of the Salpeter equation (see Ref.~\cite{RMMP}).
In order to take into account a full content of global QCD symmetries
in a systematic way, a coupled set of Dyson-Schwinger and BS equations
should be considered (see, e.g. \cite{Jain}).

Note that the number of terms ($N_{max}$) in the expansion (\ref{eq32}), which is
necessary to get stable solutions of the system of equations (\ref{eq34}),
varies with the constituent quark masses, with the value of the mixing parameter $x$,
with the state under consideration and is different for the CJ, MNK and Sal.
versions. Namely, when the quark masses increase,
$N_{max}$ decreases. The convergence of the numerical procedure used in the calculations
is better for $x \leq 0.5$ and worse for $x>0.5$. For all values of $x$
the convergence is better for the Sal. version than for the CJ and
MNK versions.

We have also calculated the mass spectrum of $q\bar{q}$ systems with the equal
quark masses from the light quark sector $(u\bar{u},s\bar{s})$.
This problem was the subject of our primary interest in the present study.
The results of calculations are shown in Figs. 4 and 5. Note
that for this sort of systems the convergence of numerical algorithm
used in the calculations
appears to be not so good for the values of the parameters
$\omega_{0}$=710 MeV and $x>0.5$. The convergence becomes
better for smaller values of $\omega_{0}$. Namely, for the $(u\bar{u})$
system for $\omega_{0}$=710 MeV  and $x=0.6$, stable solutions in the MNK version
for the $^{3}S_{1}$ state do not exist. In the CJ version such a situation holds
for other states $(^{3}P_{2},^{3}D_{1},^{3}D_{3})$ as well. For smaller values
of the potential strength parameter, e.g. $\omega_{0}$=550 MeV,
the stable solutions exist for all states
(just these results are shown in fig. 4). Further,
in this case the sequence of the energy levels corresponding to the $^{3}P_{J}$
states (the spin-orbit splitting) agrees with the experiment at $x>0.5$ in
all versions, and in the $^{3}D_{1}$ and $^{3}D_{3}$ states the agreement
appears to occur at $x<0.5$.

Consequently, on the basis of the above analysis one can conclude that
none of the 3D equations with the simple oscillator kernel considered in the
present paper, does simultaneously describe even general features of the
mass spectrum of all $q\bar q$ systems under study.

\section{Some decay characteristics of mesons}

For the investigation of the meson decay properties, the normalization
condition for the wave function  $\tilde{\Phi}_{M}(\vec p\,)= \sum\limits_
{(\alpha_{1}\alpha_{2})}\tilde{\Phi}_{M}^{(\alpha_{1}\alpha_{2})}(\vec p\,)$
is needed.
For the Salpeter wave function this condition is well known \cite{IZ}
\eq\label{eq35}
        \int\frac{d^3\vec p}{(2\pi)^{3}}\,\,\
\biggl[\mid \tilde{\Phi}_{M}^{(++)}(\vec p\,) \mid ^{2}-
\mid \tilde{\Phi}_{M}^{(--)}(\vec p\,) \mid ^{2}\biggr]=2M
\en
\noindent
As to the MW, CJ and MNK versions, the
normalization condition for the corresponding wave functions can be obtained
with the use of the fact that
the effective Green  operators (\ref{eq4}) in Eq. (\ref{eq1})
can be inverted. As a result, the corresponding full 3D Green
operators $\tilde{G}_{0eff}$ can be defined analogously to the 4D case
\eq
\tilde{G}_{eff}^{-1}(M;\vec p,\vec p~') =
(2\pi)^3\delta^{(3)}(\vec p-\vec p~')\tilde{G}_{0eff}^{-1}(M;\vec p\,)-
\hat{V}(\vec p,\vec p~')
\en
\noindent
Since $\hat{V}(\vec p,\vec p~')$ does not depend on $M$, the normalization
condition reads
\eq\label{eq37}
        \int\frac{d^3\vec p}{(2\pi)^{3}}\,\,\
\biggl[\overline{\tilde{\Phi}}_{M}(\vec p\,)=
\tilde{\Phi}_{M}^{^{+}}(\vec p\,)
\gamma_{1}^{0}\gamma_{2}^{0}\biggr]
\biggl[\frac{\partial}{\partial M}
\tilde{G}_{0eff}^{-1}(M;\vec p\,)\biggr]
\tilde{\Phi}_{M}(\vec p\,)=2M.
\en
\noindent
Using Eqs. (\ref{eq4})-(\ref{eq8}), from Eq. (\ref{eq37}) one obtains
\eq\label{eq38}
\sum\limits_{\alpha_{1}\alpha_{2}}\int\frac{d^3\vec p}{(2\pi)^{3}}
\tilde{\Phi}_{M}^{^{+}(\alpha_{1}\alpha_{2})}(\vec p\,)
f_{12}^{(\alpha_{1}\alpha_{2})}(M;p)
\tilde{\Phi}_{M}^{(\alpha_{1}\alpha_{2})}(\vec p)=2M
\en
\noindent
where
\eqa
f_{12}^{(\alpha_{1}\alpha_{2})}=\frac{\alpha_{1}E_{1}+\alpha_{2}E_{2}}{M}
\hspace*{1.2cm}\,\,\,\,\,\,\,\,\,\,\,\,\,\,\,\,\,\,\,\,\,\,\,\,\,\,\,\,\,
\mbox{(MW version)}&
\\[2mm]
f_{12}^{(\alpha_{1}\alpha_{2})}=\frac{\omega_{1}+\omega_{2}}{M}
\frac{\alpha_{1}\omega_{1}E_{2}+\alpha_{2}\omega_{2}E_{1}}
{(E_{1}+\alpha_{1}\omega_{1})(E_{2}+\alpha_{2}\omega_{2})}
\hspace*{1.2cm}
\mbox{(CJ version)}&
\\[2mm]
f_{12}^{(\alpha_{1}\alpha_{2})}=\frac{2}{D^{(\alpha_{1}\alpha_{2})}}
\biggl \{ \biggl[\frac{M}{R}B(1-y^{2})+\frac{M^{2}}{2}-2\biggl(\frac{R-M}
{2y}\biggr)^{2}\biggr]-
&\nonumber\\[2mm]
-\frac{B}{D^{(\alpha_{1}\alpha_{2})}}\biggl[\biggl(M+
\frac{\alpha_{1}\omega_{1}+\alpha_{2}\omega_{2}}{2}R-
-\frac{M^{2}}{2}-2\biggl(\frac{R-M}{2y}\biggr)^{2}+
&\nonumber\\[2mm]
+(\alpha_{1}\omega_{1}-\alpha_{2}\omega_{2})\biggl(\frac{R-M}{2y}
+\frac{M}{2}y\biggr)\biggr]\biggr \}
\hspace*{1.2cm}\,\,\,\,\,\,\,\,\,\,
\mbox{(MNK version)}&
\ena
Note that for $m_{1}=m_{2}=m$ the normalization condition (\ref{eq38})
for the MW version is reduced to (\ref{eq35}) which can be written in
the form of Eq. (\ref{eq38}) where
\eqa
f_{12}^{(\alpha_{1}\alpha_{2})}=\frac{\alpha_{1}+\alpha_{2}}{2}
\hspace*{1.2cm}
\mbox{(Sal. version)}&
\ena
From the normalization condition (\ref{eq38}) for the wave function
given by Eqs. (\ref{eq16}), (\ref{eq17}) one obtains the
normalization condition for the wave functions $R_{LSJ}^{(a1a2)}(p)$
with the use of the partial-wave expansion (\ref{eq24})
\eq\label{eq43}
\int_{0}^{\infty}\frac{p^{2} dp}{(2\pi)^{3}}
\sum\limits_{\alpha_{1}\alpha_{2}}f_{12}^{(\alpha_{1}\alpha_{2})}(M;p)
\biggl[R_{LSJ}^{(\alpha_{1}\alpha_{2})}(p)\biggr]^{2}=2M
\en
\noindent
The functions $f_{12}^{(--)}$ (40,41) have second order poles at
$p=p_{s}$, where
\eq
a(p_{s})=E_{i}^{2}-\omega_{i}^{2}(p_{s})=0,
\hspace*{1.2cm}
p_{s}=\frac{1}{2}\bigl( M^{2}+b_{0}^{2}-2(m_{1}^{2}+m_{2}^{2})\bigr)^{1/2}
\en

The functions $f_{12}^{(\pm\mp)}$ turn out to be finite
$(f_{12}^{(++)}(p_{s})$ is apparently finite). Consequently, the normalization
condition (\ref{eq43}) for the CJ and MNK versions involves a singular
integral of the type
\eq
I(x_{0})=\int_{0}^{\infty}\frac{f(x)dx}{(x-x_{0})^{2}}
\en
\noindent
which taking account of the conditions $f(0)=0=f(\infty)$ valid in our case,
can be regularized as
\eq
\int_{0}^{\infty}\frac{f(x)dx}{(x-x_{0})^{2}}=
\int_{0}^{2x_0}\frac{[f'(x)-f'(x_{0})]dx}{(x-x_{0})}+
\int_{2x_{0}}^{\infty}\frac{f'(x)dx}{x-x_{0}}
\en
\noindent
Now we can calculate the pseudoscalar ($S=L=J=0$) decay constant
$f_{P}(P \rightarrow\mu \bar{\nu})$ and the leptonic decay width of the vector
($l=0, S=J=1$) meson $\Gamma(V\rightarrow e^{-}e^{+})$ (the corresponding
decay constant is denoted by $f_{V}$). In these calculations, instead of the
functions $\tilde{\Phi}_{M}^{(\alpha_{1}\alpha_{2})}(\vec p\,)$ (\ref{eq10})
given as a column with the components
$\tilde{\Phi}_{aa}^{(\alpha_{1}\alpha_{2})}$,
$\tilde{\Phi}_{ab}^{(\alpha_{1}\alpha_{2})}$,
$\tilde{\Phi}_{ba}^{(\alpha_{1}\alpha_{2})}$,
$\tilde{\Phi}_{bb}^{(\alpha_{1}\alpha_{2})}$,
it is convenient to introduce the wave function
$\Psi^{(\alpha_{1}\alpha_{2})}$ written in the form (see Ref.~\cite{IZ})
\eq\label{eq47}
\Psi^{(\alpha_{1}\alpha_{2})}=\pmatrix{
\tilde{\Phi}_{aa}^{(\alpha_{1}\alpha_{2})}\,\,
\tilde{\Phi}_{ab}^{(\alpha_{1}\alpha_{2})}\cr
\tilde{\Phi}_{ba}^{(\alpha_{1}\alpha_{2})}\,\,
\tilde{\Phi}_{bb}^{(\alpha_{1}\alpha_{2})}}(C=i\gamma^{2}\gamma^{0})=
\pmatrix{
\tilde{\Phi}_{aa}^{(\alpha_{1}\alpha_{2})}\sigma_{y}\,\,
\tilde{\Phi}_{ab}^{(\alpha_{1}\alpha_{2})}\sigma_{y}\cr
\tilde{\Phi}_{ba}^{(\alpha_{1}\alpha_{2})}\sigma_{y}\,\,
\tilde{\Phi}_{bb}^{(\alpha_{1}\alpha_{2})}\sigma_{y}}
\en
\noindent
where $C$ is the charge conjugation operator. Then, the
decay constants $f_{P}$ and $f_{V}$ are given by the expressions
\eq\label{eq48}
\delta_{\mu 0} Mf_{P}=\sqrt{3}Tr\biggl[\Psi_{000}(\vec r=0)\gamma^{\mu}
(1-\gamma_{5})\biggr]
\en
\eq\label{eq49}
f_{V}(\lambda)=\sqrt{3}Tr\biggl[\Psi_{011\lambda}(\vec r=0)\gamma^{\mu}\biggr]
\varepsilon^{\mu}(\lambda=0,\pm 1)
\en
\noindent
Here, the factor $\sqrt{3}$ stems from the color part of the wave functions,
$\varepsilon^{\mu}(\lambda)$ is the polarization vector of the meson and
\eq\label{eq50}
\Psi_{LSJM_{J}}(\vec r=0)=\int\frac{d^3\vec p}{(2\pi)^{3}}\,
\biggl[\Psi_{LSJM_{J}}(\vec p\,)=\sum\limits_{\alpha_{1}\alpha_{2}}
\Psi_{LSJM_{J}}^{(\alpha_{1}\alpha_{2})}(\vec p\,)\biggr]
\en
\noindent
Using Eqs. (\ref{eq16}), (\ref{eq17}), (\ref{eq24}),
(\ref{eq47}) and (\ref{eq50}), from Eqs. (\ref{eq48}) and (\ref{eq49})
one obtains
\eq\label{eq51}
f_{P}=\frac{\sqrt{24\pi}}{M}\int_{0}^{\infty}\frac{p^{2} dp}{(2\pi)^{3}}
\sum\limits_{\alpha_{1}\alpha_{2}}\biggl[N_ {1}^{(\alpha_{1})}(p)
N_ {2}^{(\alpha_{2})}(p)-\alpha_{1}\alpha_{2}N_ {1}^{(-\alpha_{1})}(p)
N_ {2}^{(-\alpha_{2})}(p)\biggr]R_{000}^{(\alpha_{1}\alpha_{2})}(p)
\en
\eqa
f_{V}(\lambda)&=&\biggl \{-\sqrt{24\pi}\int_{0}^{\infty}\frac{p^{2}dp}
{(2\pi)^{3}}\sum\limits_{\alpha_{1}\alpha_{2}}
\biggl[N_ {1}^{(\alpha_{1})}(p)N_ {2}^{(\alpha_{2})}(p)+
\nonumber\\[2mm]
&+&\frac{\alpha_{1}\alpha_{2}}{3}N_ {1}^{(-\alpha_{1})}(p)
N_ {2}^{(-\alpha_{2})}(p)\biggr]R_{011}^{(\alpha_{1}\alpha_{2})}(p)
\biggr \}\delta_{\lambda 0}
\label{eq52}
\ena
\noindent
For a given $f_{V}$ (\ref{eq52}) the leptonic decay width of the vector meson
is given by
\eq\label{eq53}
\Gamma(V \rightarrow e^{-} e^{+})=\frac{\alpha^{2}_{eff}}{4 \pi M^{3}}
\frac{1}{3}\sum\limits_{\lambda=0,\pm}\mid f_{V}(\lambda) \mid^{2}
\nonumber\en
where
$$
\alpha^{2}_{eff}=\frac{1}{137}\biggl(\frac{1}{2},\frac{1}{18},
\frac{1}{9} \biggr)
$$
\noindent
for $\varrho^{0}, \omega$ and $\varphi$ mesons, respectively.

Finally, using Eqs. (\ref{eq32}), (\ref{eq33}), (\ref{eq43}),
(\ref{eq51}), (\ref{eq52}) and (\ref{eq53}), we obtain
\eqa
f_{P}&=&\frac{\sqrt{6}p_{0}^{\frac{3}{2}}}{\pi\sqrt{M}} \mid
\sum\limits_{\alpha_{1}\alpha_{2}}\int_{0}^{\infty}z^{2} dz
\biggl[N_ {1}^{(\alpha_{1})}(p_{0},z)N_ {2}^{(\alpha_{2})}(p_{0},z)-
\nonumber\\[2mm]
&-&\alpha_{1}\alpha_{2}N_ {1}^{(-\alpha_{1})}(p_{0},z)
N_ {2}^{(-\alpha_{2})}(p_{0},z)\biggr]\bar R_{000}^{(\alpha_{1}\alpha_{2})}(z)
\mid
\label{eq55}
\ena
\noindent
\eqa
\Gamma(V \rightarrow e^{-} e^{+})&=&
\frac{4\alpha_{eff}^{2}p_{0}}{(2\pi)^{3}M^{2}}
\mid \sum\limits_{\alpha_{1}\alpha_{2}}\int_{0}^{\infty}z^{2}dz
\biggl[N_ {1}^{(\alpha_{1})}(p_{0},z)N_ {2}^{(\alpha_{2})}(p_{0},z)+
\nonumber\\[2mm]
&+&\frac{\alpha_{1}\alpha_{2}}{3}N_ {1}^{(-\alpha_{1})}(p_{0},z)
N_ {2}^{(-\alpha_{2})}(p_{0},z)\biggr]\bar R_{011}^{(\alpha_{1}\alpha_{2})}(z)
\mid ^{2}
\label{eq56}
\ena
\noindent
where the functions
\eq
\bar R_{LSJ}^{(\alpha_{1}\alpha_{2})}(z)=\sum\limits_{n=0}^{\infty}
C_{LSJ}(\alpha_{1}\alpha_{2})\bar R_{nL}(z)
\en
\noindent
satisfy the normalization condition
\eq
\sum\limits_{\alpha_{1}\alpha_{2}}\int_{0}^{\infty}z^{2}dz
f^{(\alpha_{1}\alpha_{2})}_{12}(M;p_{0},z)\biggl[
\bar R_{LSJ}^{(\alpha_{1}\alpha_{2})}(z)\biggr]^{2}=1
\en
\noindent
The results of numerical calculations of the quantities $f_{P}$ and
$\Gamma$ defined by Eqs. (\ref{eq55}) and (\ref{eq56}), are given in Table I.
We see from Table I that the calculated values of $f_{P}$
in the MNK and CJ versions, as a rule, are smaller than in the Sal. version
and this fact is related to the presence of the contributions from
the "$+-$"  and "$-+$" components of the wave function (contribution from
the "$--$" component is negligibly small). Further, the calculated
value of $f_P$ is larger in the CJ version than in the MNK version.
The calculated value of the quantity $\Gamma(V \rightarrow e^{-} e^{+})$
weakly depends on the particular choice of the 3D reduction scheme of
the BS equation. With the increase of the mixing parameter $x$ both the
quantities $f_{P}$ and $\Gamma(V \rightarrow e^{-} e^{+})$ slightly increase.
The calculated values of $f_{P}$ and $\Gamma(V \rightarrow e^{-} e^{+})$
are smaller than the experimental ones.
\vspace*{.6cm}

On the basis of the analysis of the different versions of the 3D
reductions of the bound state BS equation carried out in the
present paper, one arrives at the following conclusions:

The existence/nonexistence of stable solutions of the 3D bound-state
equations critically depends on the value of the scalar-vector mixing parameter
$x$. For all 3D versions (Sal, MNK, CJ) stable solutions coexist
for the value of $x$ from a rather restricted interval
$0.3\leq x \leq 0.6-0.7$. The level ordering
in the mass spectrum is similar for all versions under consideration.
However, the sequence of the calculated energy levels agrees with the
experiment for some states at $ x < 0.5$ and for other states at
$ x > 0.5$.
Consequently, a simultaneous description of even general features of
the meson whole mass spectrum turns out not to be possible for a given
value of $x$ from the above-mentioned interval.
It is interesting to investigate the dependence of this results on
the form of the confining potential. Also, it is interesting
to study how it changes when the one-gluon exchange potential is taken
into account. This aspect of the problem will be considered at the next
step of our investigation. Further, we plan to include the
't Hooft effective interaction in our potential in order
to (phenomenologically) take into account the effect of spontaneous
breaking of chiral symmetry which is important in the pseudoscalar
($^{1}S_{0}$) sector of the constituent model.

The calculated leptonic decay characteristics of mesons are quite
insensitive to the particular 3D reduction scheme used and
give an acceptable description of experimental data.
In future, we also plan to study the validity of this conclusion
for a wider class of realistic interquark potentials.

{\it Acknowledgments.} One of the authors (A.R.) acknowledges the financial
support from the Russian Foundation for Basic Research under contract
96-02-17435-a.

\newpage

\centerline{\large{\bf TABLE CAPTIONS}}
\vspace*{.3cm}

\noindent {\bf Table I}\\
The pseudoscalar decay constant
$f_{P}(P \rightarrow\mu \bar{\nu})$ (in MeV) and the leptonic decay
width $\Gamma (V \rightarrow e^{-} e^{+})$ (in KeV) with the allowance for
only (++) and all components of the wave function.

\vspace*{2.cm}

\centerline{\large{\bf FIGURE CAPTIONS}}

\vspace*{.3cm}

\noindent {\bf Fig. 1}\\
The mass spectrum (in GeV) of the $u\bar s$ system for the different
3D equations and different values of the scalar-vector mixing parameter
$x$. The multiplicity of degenerate levels is indicated by the number
inside the dash ($\omega_0=710~MeV$).

\vspace*{.3cm}

\noindent {\bf Fig. 2}\\
The same as in Fig. 1 for the $c\bar u$ system.

\vspace*{.3cm}

\noindent {\bf Fig. 3}\\
The same as in Fig. 1 for the $c\bar s$ system.

\vspace*{.3cm}

\noindent {\bf Fig. 4}\\
The mass spectrum (in GeV) of the $u\bar u$ system for the different
3D equations and different values of the scalar-vector mixing parameter
$x$. The multiplicity of degenerate levels is indicated by the number
inside the dash. The parameter $\omega_0$ for the $^{3}S_{1}$ state in
the MNK version and for all states in the
CJ version at $x=0.6$ is fixed at the value $550~MeV$, otherwise
$\omega_0=710~MeV$.

\vspace*{.3cm}

\noindent {\bf Fig. 5}\\
The same as in Fig. 1 for the $s\bar s$ system.

\newpage

\noindent {\bf Table I}

\vspace*{1.cm}

\begin{tabular}{|c|r|c|c|c|c|c|c|c|c|}
\hline
\multicolumn{2}{|c|}{~Decay characteristics~}&
\multicolumn{4}{c|}{$f_{P}(P \rightarrow\mu \bar{\nu})$}&
\multicolumn{4}{c|}{$\Gamma(V\rightarrow e^{-}e^{+})$}\cr
\hline
\multicolumn{2}{|c|}{Meson}&
\multicolumn{2}{c|}{$D(c\bar{d})$}&
\multicolumn{2}{c|}{$D_s(c\bar{s})$}&
\multicolumn{2}{c|}{$\varrho(u\bar{u})$}&
\multicolumn{2}{c|}{$\varphi(s\bar{s})$}\cr
\hline
\multicolumn{1}{|c|}{~Versions~              }&
\multicolumn{1}{c|}{$\alpha_{1}\alpha_{2}$}&
\multicolumn{1}{c|}{~$x$=0.3~               }&
\multicolumn{1}{c|}{~$x$=0.5~               }&
\multicolumn{1}{c|}{~$x$=0.3~               }&
\multicolumn{1}{c|}{~$x$=0.5~               }&
\multicolumn{1}{c|}{~$x$=0.3~               }&
\multicolumn{1}{c|}{~$x$=0.5~               }&
\multicolumn{1}{c|}{~$x$=0.3~               }&
\multicolumn{1}{c|}{~$x$=0.5~               }\cr
\hline              
Sal&\multicolumn{1}{c|}{++}&
\multicolumn{1}{c|}{149.}&
\multicolumn{1}{c|}{156.}&
\multicolumn{1}{c|}{177.}&
\multicolumn{1}{c|}{183.}&
\multicolumn{1}{c|}{3.94}&
\multicolumn{1}{c|}{4.31}&
\multicolumn{1}{c|}{0.84}&
\multicolumn{1}{c|}{0.90}\\
"&\multicolumn{1}{c|}{all}&
\multicolumn{1}{c|}{148.}&
\multicolumn{1}{c|}{155.}&
\multicolumn{1}{c|}{176.}&
\multicolumn{1}{c|}{182.}&
\multicolumn{1}{c|}{3.96}&
\multicolumn{1}{c|}{4.38}&
\multicolumn{1}{c|}{0.84}&
\multicolumn{1}{c|}{0.90}\\
\hline
MNK&\multicolumn{1}{c|}{++}&
\multicolumn{1}{c|}{148.}&
\multicolumn{1}{c|}{155.}&
\multicolumn{1}{c|}{176.}&
\multicolumn{1}{c|}{181.}&
\multicolumn{1}{c|}{4.44}&
\multicolumn{1}{c|}{4.85}&
\multicolumn{1}{c|}{0.89}&
\multicolumn{1}{c|}{0.95}\\
"&\multicolumn{1}{c|}{all}&
\multicolumn{1}{c|}{141.}&
\multicolumn{1}{c|}{145.}&
\multicolumn{1}{c|}{168.}&
\multicolumn{1}{c|}{172.}&
\multicolumn{1}{c|}{4.11}&
\multicolumn{1}{c|}{4.20}&
\multicolumn{1}{c|}{0.85}&
\multicolumn{1}{c|}{0.89}\\
\hline
CJ&\multicolumn{1}{c|}{++}&
\multicolumn{1}{c|}{150.}&
\multicolumn{1}{c|}{158.}&
\multicolumn{1}{c|}{179.}&
\multicolumn{1}{c|}{186.}&
\multicolumn{1}{c|}{3.98}&
\multicolumn{1}{c|}{4.45}&
\multicolumn{1}{c|}{0.86}&
\multicolumn{1}{c|}{0.94}\\
"&\multicolumn{1}{c|}{all}&
\multicolumn{1}{c|}{149.}&
\multicolumn{1}{c|}{152.}&
\multicolumn{1}{c|}{176.}&
\multicolumn{1}{c|}{180.}&
\multicolumn{1}{c|}{3.91}&
\multicolumn{1}{c|}{4.23}&
\multicolumn{1}{c|}{0.85}&
\multicolumn{1}{c|}{0.90}\\
\hline
\multicolumn{2}{|c|}{Expt.}&
\multicolumn{2}{c|}{$<$ 220 }&
\multicolumn{2}{c|}{170 $\div$ 180}&
\multicolumn{2}{c|}{6.8 $\pm$0.3}&
\multicolumn{2}{c|}{1.37 $\pm$ 0.05}\cr
\hline
\end{tabular}

\newpage

\begin{figure}
\vspace*{-1.cm}\hspace*{-2.cm}
{\bf
\mbox{\epsfysize=23cm\epsffile{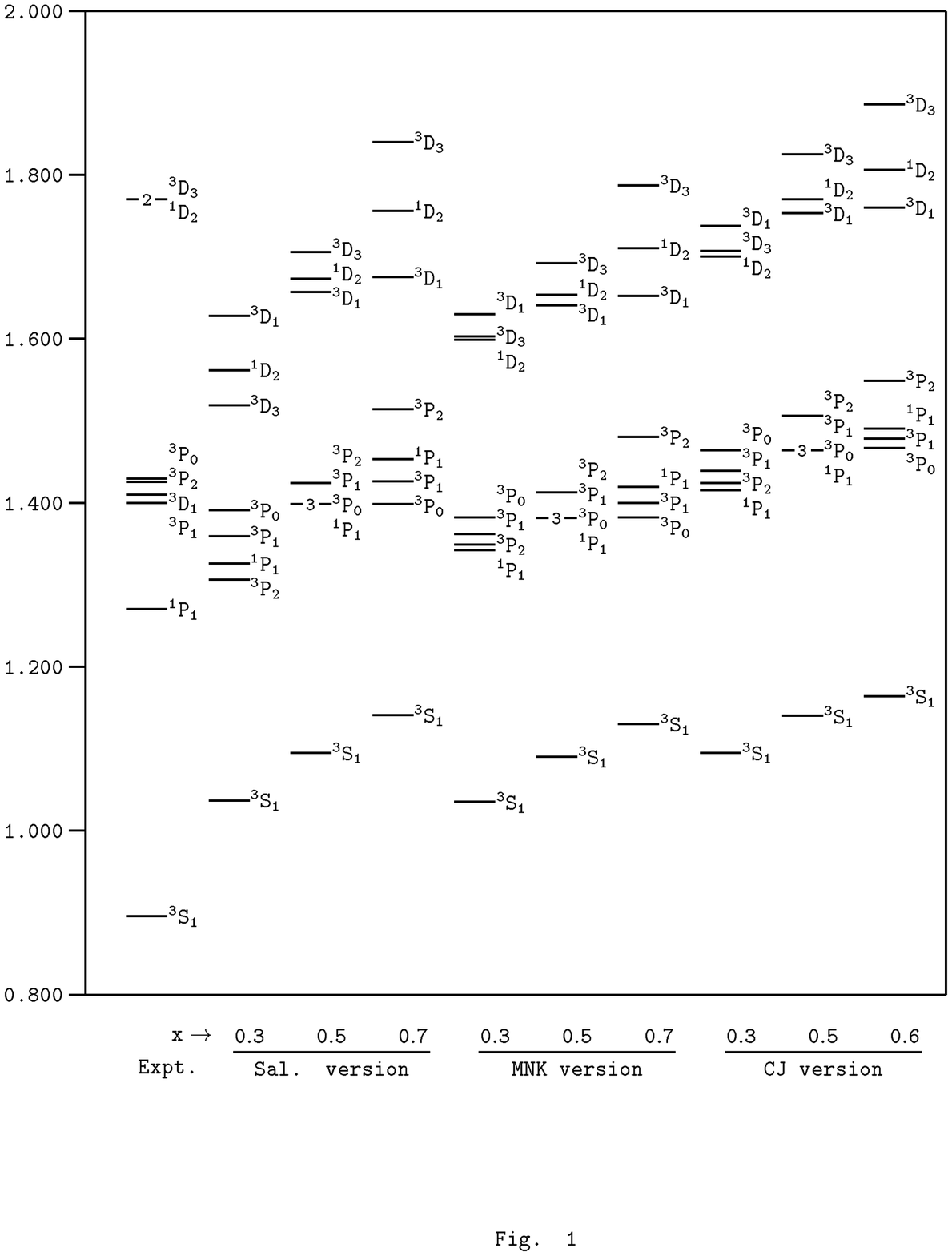}}}
\end{figure}

\newpage
\begin{figure}
\vspace*{-2.cm}\hspace*{-2.cm}
{\bf
\mbox{\epsfysize=23cm\epsffile{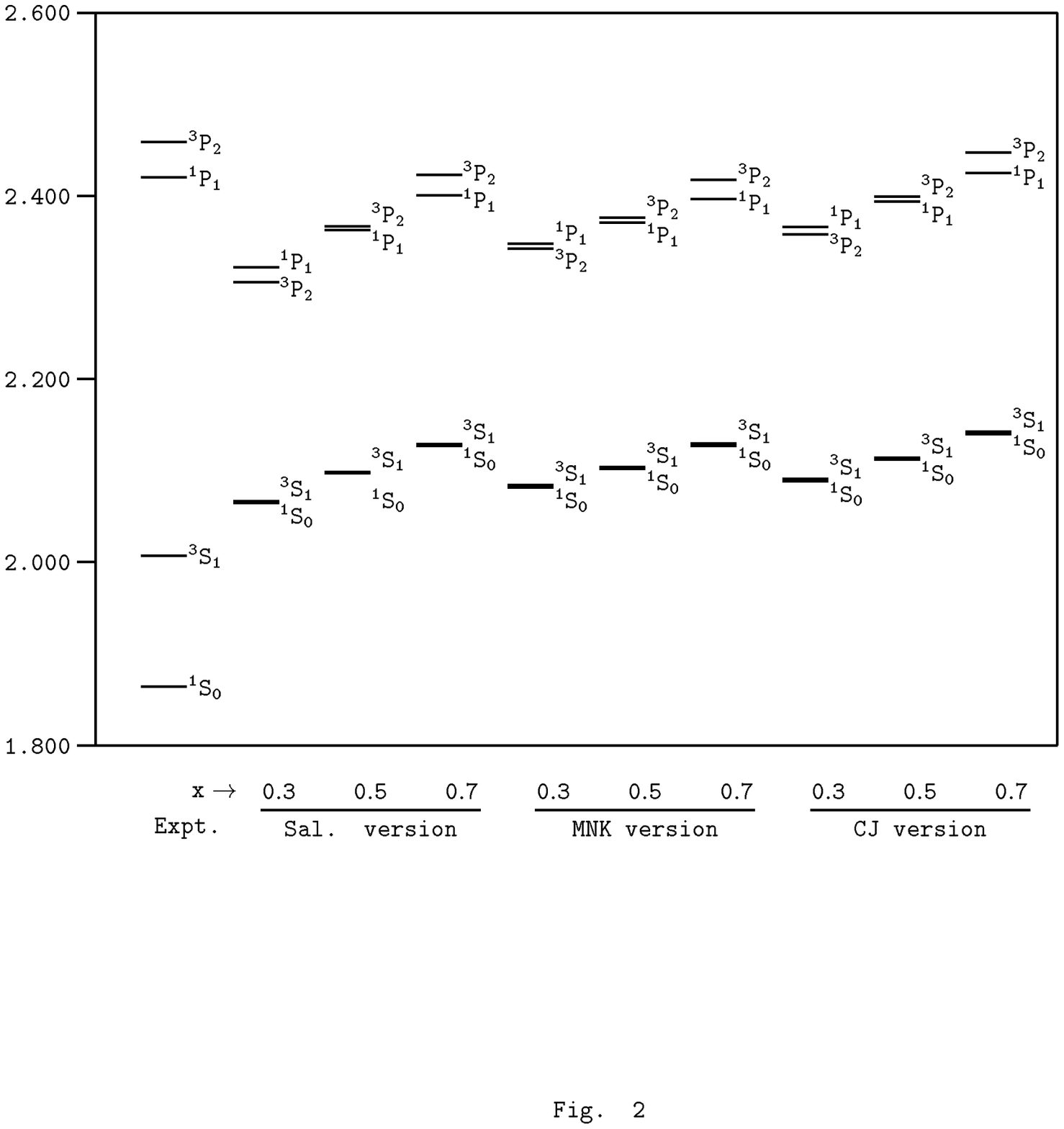}}}
\end{figure}

\newpage
\begin{figure}
\vspace*{-2.cm}\hspace*{-2.cm}
{\bf
\mbox{\epsfysize=23cm\epsffile{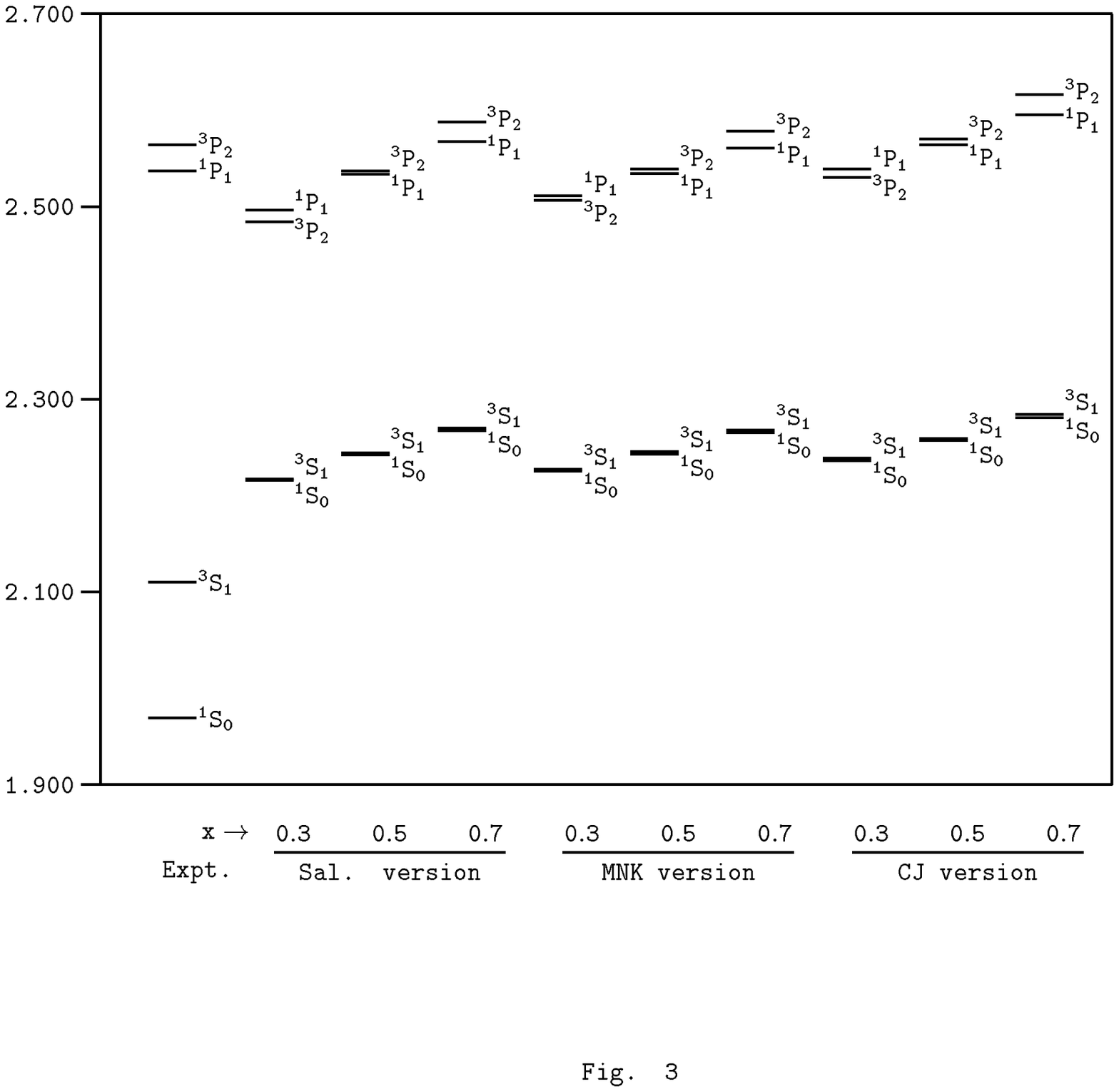}}}
\end{figure}

\newpage
\begin{figure}
\vspace*{-2.cm}\hspace*{-2.cm}
{\bf
\mbox{\epsfysize=23cm\epsffile{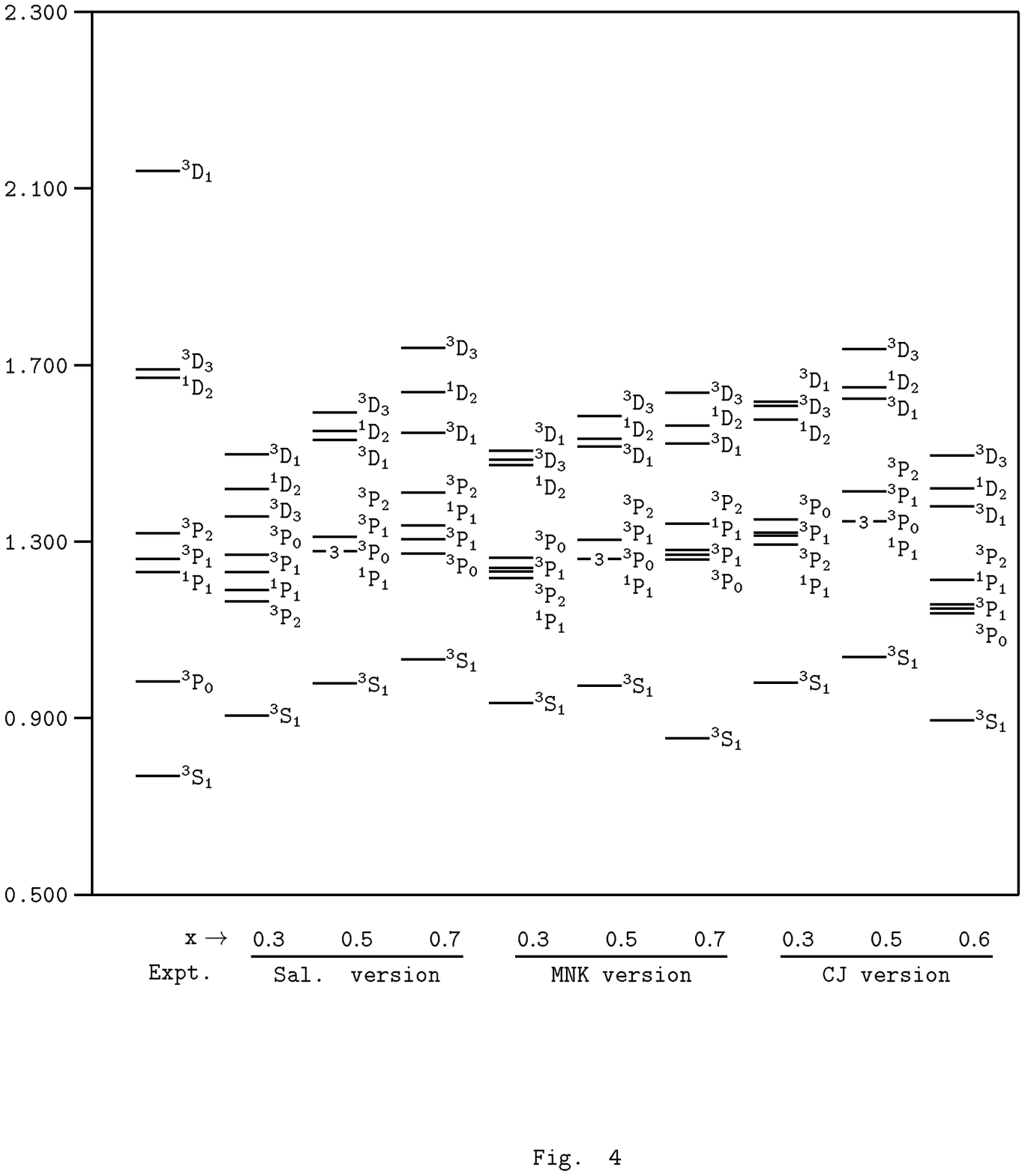}}}
\end{figure}

\newpage
\begin{figure}
\vspace*{-2.cm}\hspace*{-2.cm}
{\bf
\mbox{\epsfysize=23cm\epsffile{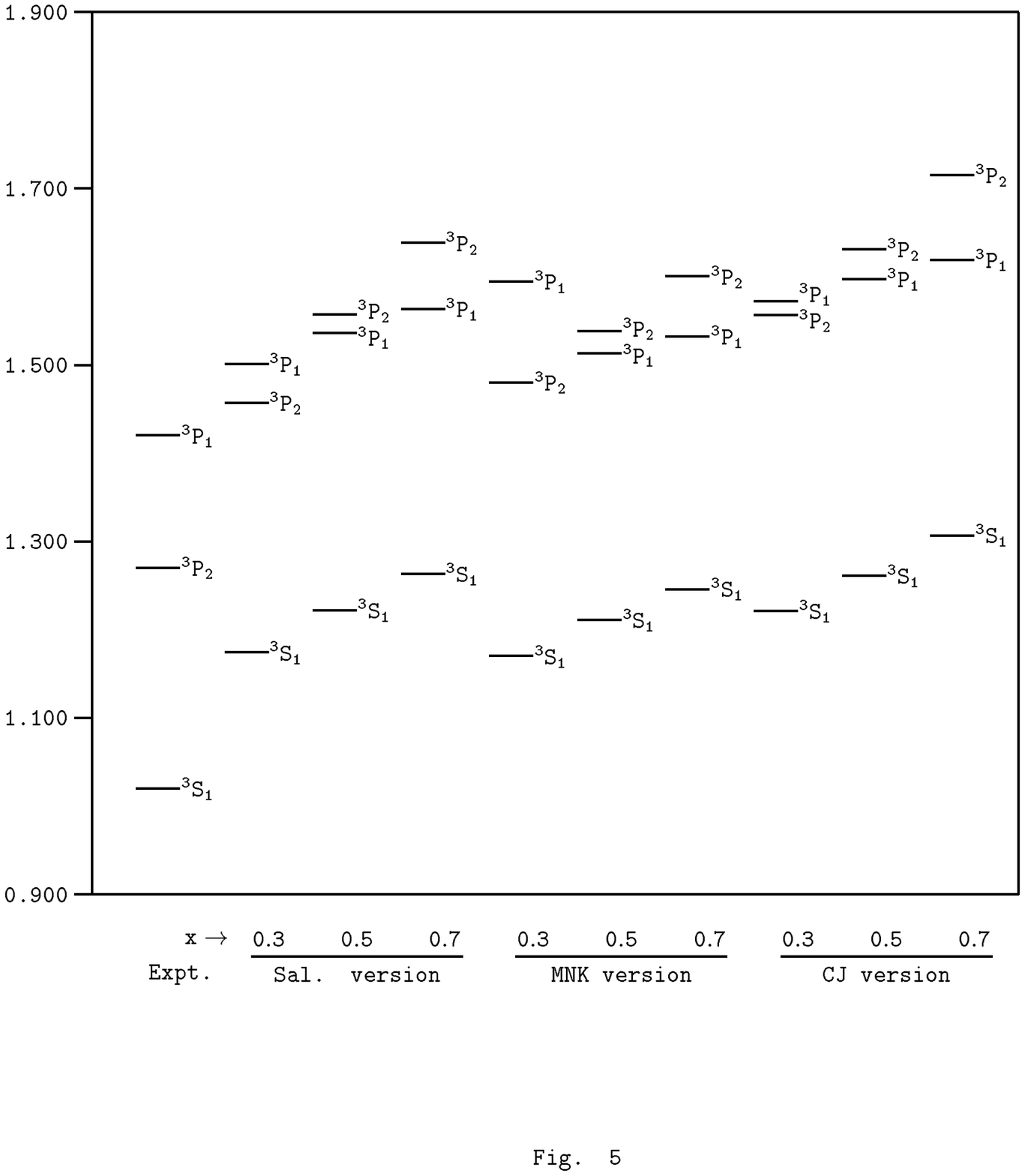}}}
\end{figure}
\end{document}